\documentclass[manuscript]{acmart}

\usepackage{amsmath}
\usepackage{algorithmic}
\usepackage{graphicx}
\usepackage{textcomp}
\usepackage{xcolor}
\usepackage{colortbl}
\usepackage{enumerate}
\usepackage{multirow}
\usepackage{dirtytalk}

\AtBeginDocument{%
  }

\setcopyright{None}

\renewcommand\footnotetextcopyrightpermission[1]{} 
    
\begin{document}

\title{Applications of Positive Unlabeled (PU) and Negative Unlabeled (NU) Learning in Cybersecurity}

\author{Robert Dilworth}
\affiliation{
  \institution{Mississippi State University}
  \city{Starkville}
  \state{Mississippi}
  \country{USA}
}
\email{rkd103@msstate.edu}

\author{Charan Gudla}
\affiliation{
  \institution{Mississippi State University}
  \city{Starkville}
  \state{Mississippi}
  \country{USA}
}
\email{gudla@cse.msstate.edu}

\begin{abstract}
    This paper explores the relatively underexplored application of Positive Unlabeled (PU) Learning and Negative Unlabeled (NU) Learning in the cybersecurity domain. While these semi-supervised learning methods have been applied successfully in fields like medicine and marketing, their potential in cybersecurity remains largely untapped. The paper identifies key areas of cybersecurity--such as intrusion detection, vulnerability management, malware detection, and threat intelligence--where PU/NU learning can offer significant improvements, particularly in scenarios with imbalanced or limited labeled data. We provide a detailed problem formulation for each subfield, supported by mathematical reasoning, and highlight the specific challenges and research gaps in scaling these methods to real-time systems, addressing class imbalance, and adapting to evolving threats. Finally, we propose future directions to advance the integration of PU/NU learning in cybersecurity, offering solutions that can better detect, manage, and mitigate emerging cyber threats.
\end{abstract}


\maketitle

\pagestyle{plain}

\section{Introduction}

    Cybersecurity is a critical field, tasked with protecting digital assets from attacks. Traditional supervised learning methods require both positive (attack) and negative (non-attack) labels for building robust models. However, in many cybersecurity scenarios, negative labels are difficult to acquire, or the data is heavily skewed towards positive examples. Positive Unlabeled (PU) Learning and Negative Unlabeled (NU) Learning offer a way to address these challenges by using limited labeled data and a large pool of unlabeled data. This paper explores how these semi-supervised learning techniques can be applied to various subfields of cybersecurity, from intrusion detection to threat hunting.

\section{Contributions of the Body of Work}

    The primary contribution of this paper is the systematic identification and exploration of PU and NU Learning applications in cybersecurity. This work:
    \begin{itemize}
        \item Identifies subfields within cybersecurity that could benefit from PU and NU Learning.
        \item Provides a theoretical framework to justify the applicability of these methods.
        \item Discusses existing research and identifies key gaps in the literature.
        \item Proposes potential future research directions in these areas.
    \end{itemize}

\section{Research Questions}

    The research addresses the following key questions:
    \begin{itemize}
        \item In which subfields of cybersecurity can PU and NU Learning provide tangible improvements over traditional implementations?
        \item How can PU and NU Learning be conceptually justified in the context of cybersecurity?
        \item What are the limitations and challenges of applying these methods to cybersecurity datasets?
    \end{itemize}

\section{Paper Structure}

    \tableofcontents

\section{Background}

    \subsection{Positive Unlabeled Learning}

        PU Learning involves training a binary classifier using only positive and unlabeled data. The goal is to develop a model that distinguishes between positive (attack) and negative (non-attack) classes, with the negative class largely hidden in the unlabeled data. Mathematically, this can be modeled as minimizing the classification risk:
        \begin{equation}
        R(f) = \mathbb{E}_{x \in P} [\ell(f(x), 1)] + \mathbb{E}_{x \in U} [\ell(f(x), -1)]
        \end{equation}
        where \( P \) is the set of positive examples, \( U \) is the unlabeled set, and \( \ell \) is the loss function.
    
    \subsection{Negative Unlabeled Learning}
    
        In NU Learning, the negative class is known, and the goal is to identify the positive class within the unlabeled data. This is useful in scenarios like anomaly detection, where the majority class is non-anomalous, but anomalies (positive class) need to be identified from unlabeled data. This can be formulated as:
        \begin{equation}
        R(f) = \mathbb{E}_{x \in N} [\ell(f(x), -1)] + \mathbb{E}_{x \in U} [\ell(f(x), 1)]
        \end{equation}
        where \( N \) is the negative (benign) set, and \( U \) is the unlabeled set.

    \subsection{Cybersecurity}
    
        Cybersecurity encompasses a wide range of subfields, each focusing on protecting various aspects of digital systems and information. As the digital landscape becomes increasingly complex, so do the types of threats, requiring specialized approaches to mitigate potential attacks. Positive Unlabeled (PU) and Negative Unlabeled (NU) Learning offer new possibilities for solving classification problems where labeled data is scarce, and the majority of data is unlabeled. This section provides an overview of the major subfields of cybersecurity and discusses the potential for applying PU and NU Learning to these areas.
        
        \subsubsection{Cryptography}
        
            Cryptography involves the design and analysis of secure communication protocols and algorithms, with the primary goals of ensuring confidentiality, integrity, and authentication of data. In formal terms, cryptographic systems rely on transformations \( T_k \), where \( k \) is a cryptographic key, that take plaintext data \( P \) and convert it into ciphertext \( C \):
            \begin{equation}
            C = T_k(P)
            \end{equation}
            Decryption reverses this process:
            \begin{equation}
            P = T_k^{-1}(C)
            \end{equation}
            The security of cryptographic systems depends on the intractability of discovering \( P \) without access to the key \( k \). PU learning could be applied to anomaly detection in cryptographic usage, identifying irregular patterns in encrypted communication to detect unauthorized decryption attempts or cryptographic key misuse.
        
        \subsubsection{Network Security}
        
            Network security focuses on protecting computer networks from attacks, ensuring data transmitted across the network remains secure. This typically involves implementing firewalls, Intrusion Detection Systems (IDS), and secure network protocols. Network traffic can be represented as sequences of packets \( \{p_i\} \), each containing information about source, destination, and data payload:
            \begin{equation}
            p_i = (s_i, d_i, D_i)
            \end{equation}
            where \( s_i \) and \( d_i \) represent the source and destination, and \( D_i \) is the data. Network intrusion detection can be framed as a PU learning problem, where known malicious traffic forms the positive class, and all other traffic remains unlabeled. NU learning could also be applied to label benign traffic and identify anomalies.
        
        \subsubsection{Application Security}
        
            Application security involves the identification and mitigation of vulnerabilities in software applications to prevent unauthorized access, data breaches, and other cyberattacks. Each application \( A \) has a set of potential vulnerabilities \( \{V_1, V_2, \dots, V_n\} \), where the goal is to detect and mitigate each \( V_i \). PU learning could be used to detect unknown vulnerabilities by learning from a small set of known positive vulnerabilities and treating all other software behaviors as unlabeled data. This could significantly improve vulnerability management by focusing on code patterns that resemble known flaws.
            
        \subsubsection{Incident Response and Forensics}
            
            Incident response deals with detecting, analyzing, and responding to security incidents, while forensics focuses on collecting and preserving digital evidence for legal investigations. Formally, incident logs can be denoted as a sequence of events \( E = \{e_1, e_2, \dots, e_n\} \), where each \( e_i \) contains information about timestamps, users, and actions:
            \begin{equation}
            e_i = (t_i, u_i, a_i)
            \end{equation}
            Given the rarity of security breaches, incident response could benefit from NU learning by focusing on identifying positive instances of attacks within a dataset dominated by negative (normal) activity logs. Forensics, similarly, could use PU learning to identify anomalies in system behavior logs for proactive evidence gathering.
        
        \subsubsection{Risk Management}
        
            Risk management in cybersecurity involves identifying, assessing, and mitigating potential threats to an organization’s assets, operations, and reputation. The risk level \( R \) for a given asset can be modeled as a function of threat likelihood \( L(T) \), impact \( I \), and vulnerability \( V \):
            \begin{equation}
            R = L(T) \times I \times V
            \end{equation}
            In practice, risk management processes are data-intensive and can benefit from PU learning to classify unknown risks based on patterns identified in previously assessed threats, while NU learning can identify unknown negative risks based on minimal positive samples.
        
        \subsubsection{Security Operations (SIEM and SOAR)}
        
            Security operations involve real-time monitoring and response to security events through systems like Security Information and Event Management (SIEM) and Security Orchestration, Automation, and Response (SOAR). Let \( A = \{a_1, a_2, \dots, a_n\} \) represent the actions taken in response to detected threats \( T = \{t_1, t_2, \dots, t_m\} \). Real-time data monitoring generates alerts based on predefined threat patterns, but false positives are common. PU learning can refine SIEM alerts by classifying which events are actual threats, using a small set of labeled alerts as the positive class.
            
        \subsubsection{Identity and Access Management (IAM)}
            
            Identity and Access Management (IAM) controls who has access to systems, applications, and data, typically through user roles and privileges. Access decisions are based on a user’s identity \( I \), their role \( R \), and the resource \( S \) being requested:
            \begin{equation}
            A(I, R, S) = \text{Grant or Deny}
            \end{equation}
            IAM systems generate massive amounts of access logs, which could benefit from PU learning to detect abnormal access requests or privilege escalations. NU learning could similarly identify rare, malicious privilege changes by learning from typical access patterns.
        
        \subsubsection{Security Architecture and Engineering}
        
            This subfield involves designing and maintaining secure systems and infrastructures. The architecture of a system \( \mathcal{S} \) is defined by its components \( C = \{C_1, C_2, \dots, C_n\} \), and their interactions \( I \), which must be secured against threats. PU learning can help identify potential weaknesses in system architectures by learning from a small set of known vulnerabilities while treating other system components and interactions as unlabeled data.
            
        \subsubsection{Vulnerability Management}
            
            Vulnerability management involves identifying, assessing, and remediating vulnerabilities in software, systems, and networks. Vulnerability \( V_i \) in a system can be modeled as a function of its exploitability \( E(V_i) \) and impact \( I(V_i) \):
            \begin{equation}
            V_i = E(V_i) \times I(V_i)
            \end{equation}
            PU learning can be used to predict vulnerabilities that have not yet been exploited by using patterns from a small set of known vulnerabilities. NU learning could similarly identify software areas unlikely to have vulnerabilities by learning from normal system operation logs.
        
        \subsubsection{Threat Intelligence}
        
            Threat intelligence focuses on gathering, analyzing, and disseminating information about cyber threats, adversaries, and their tactics, techniques, and procedures (TTPs). PU learning can be used to classify new threat indicators from a dataset of known threat behaviors, while NU learning could assist in identifying emerging threats that are not yet fully understood.
        
        \subsubsection{Compliance and Regulatory Requirements}
        
            Compliance ensures that cybersecurity practices meet the necessary laws, regulations, and industry standards. Compliance reports and logs \( L = \{l_1, l_2, \dots, l_n\} \) must be audited regularly. PU learning could identify non-compliant activities from a set of previously labeled incidents of non-compliance, while NU learning could be used to confirm which activities are fully compliant without requiring exhaustive labeling.
        
        \subsubsection{Cyber Threat Hunting}
        
            Cyber threat hunting involves proactively searching for advanced persistent threats (APTs) that may have evaded traditional detection methods. Threat hunting systems can leverage PU learning to classify suspicious activity as potential APT behavior based on minimal known attacks. NU learning can assist in identifying benign traffic patterns from the vast quantities of unlabeled data generated by network activity.
        
        \subsubsection{Penetration Testing and Red Teaming}
        
            Penetration testing and red teaming simulate real-world cyberattacks to assess an organization's security posture. The outcomes of these tests, such as successful exploits \( E \), can be modeled using PU learning to identify unknown attack vectors, treating simulated attack successes as the positive class and failed attempts as unlabeled data. This allows for identifying new vulnerabilities that were not directly targeted during testing.
        
        \subsubsection{Security Awareness and Training}
        
            Security awareness and training are critical for educating employees on cybersecurity best practices. Employee behavior during training sessions \( B = \{b_1, b_2, \dots, b_n\} \) can be monitored, and PU learning can be applied to classify risky behaviors, with known security violations as the positive class. This allows organizations to focus on individuals or departments that need additional training.
        
        \subsubsection{Internet of Things (IoT) Security}
        
            The Internet of Things (IoT) introduces new security challenges due to the proliferation of connected devices. IoT devices generate a large amount of unlabeled data, and PU learning can be applied to classify abnormal device behavior, using a small set of known device vulnerabilities as the positive class. NU learning could also be applied to identify benign device behaviors, helping to isolate anomalous activities that could indicate compromised devices.

\section{Literature Review}

    Positive Unlabeled (PU) and Negative Unlabeled (NU) Learning have garnered increasing attention in various fields, particularly in domains where labeled data is scarce. These methods have demonstrated potential in addressing the challenges posed by imbalanced datasets, where the majority of samples remain unlabeled, and labeled samples are limited to a single class, typically the positive class. In cybersecurity, where attacks are often rare events and negative examples are difficult to label, these semi-supervised learning techniques provide a promising solution. This section reviews the existing literature on PU and NU Learning, focusing on their applications and potential in cybersecurity.
    
    \subsection{PU Learning in Cybersecurity}
        The concept of PU learning has been explored across various applications in cybersecurity, notably in areas such as network intrusion detection, email phishing detection, and malware detection. PU learning allows the use of a small set of positive examples and a large pool of unlabeled examples to train classifiers, which is particularly useful in cybersecurity, where acquiring a large set of labeled negative samples is challenging.
    
        \textit{Jaskie et al.} \cite{Jaskie2019} discuss the limited representation of PU learning in security applications, despite its growing use in other fields like medical and business analytics. They suggest that PU learning could be highly beneficial for network intrusion detection, where identifying attacks from a large pool of normal traffic is essential. The authors also highlight applications such as cyber-attack detection on IoT platforms, where existing attacks can form the positive class, and other data points remain unlabeled, making it an ideal case for PU learning.
        
        \textit{Wang et al.} \cite{Wang2024} introduce a PU-based approach for detecting Cross-Site Scripting (XSS) attacks, one of the most prevalent web security threats. Traditional detection systems, such as Web Application Firewalls (WAF) or Intrusion Detection Systems (IDS), often generate many false positives due to blind scanning by attackers. The PU learning model proposed by Wang et al. mitigates this issue by training on responses from web servers rather than relying solely on HTTP requests, offering a more nuanced detection of successful XSS attacks.
        
        In the context of email security, \textit{Qachfar et al.} \cite{Qachfar2022} applied PU learning to the detection of phishing emails. Their approach leverages a semi-supervised technique that uses bootstrap aggregating (bagging) to label unlabeled email samples. By probabilistically labeling these samples, the model increases its generalization ability and improves the detection of phishing attacks, which are a growing concern for both individuals and organizations.
    
    \subsection{NU Learning in Cybersecurity}
        NU Learning, while similar in concept to PU Learning, operates in situations where only negative samples are labeled, and the task is to identify positive examples within the unlabeled data. This method is particularly useful in anomaly detection tasks, where the majority class is negative (normal behavior), and the goal is to detect rare anomalies (attacks).
        
        \textit{Ding et al.} \cite{Ding2022} introduced NURD, a novel NU learning model for predicting stragglers in computer systems. The model trains on negative and unlabeled data, predicting straggling tasks to trigger proactive interventions. The same principles can be applied to cybersecurity, where detecting anomalies in system performance could signal potential security breaches. Ding et al. highlight that NU learning is particularly well-suited for such real-time predictions, as it does not require positive examples to function effectively.
        
        In the realm of medical anomaly detection, \textit{Swazinna et al.} \cite{Swazinna2019} applied NU learning to the detection of Multiple Sclerosis (MS) lesions using MRI scans. While this is outside cybersecurity, their methodology of using NU learning to distinguish between healthy and abnormal data from patient scans mirrors potential cybersecurity applications, such as detecting malware from a largely benign dataset.
        
        \textit{Choi et al.} \cite{Choi2023} extended NU learning to scenarios where normal data can be labeled, and the task is to identify anomalies from unlabeled data. Their iterative training procedure incrementally labels the most likely normal samples from the unlabeled set, which is analogous to identifying safe network traffic before isolating anomalies that could indicate cyber threats.
    
    \subsection{Applications of PU and NU Learning in Vulnerability Detection and Malware Identification}
        The detection of software vulnerabilities is another area where PU learning has shown promise. \textit{Wen et al.} \cite{Wen2023} developed a PU-based model named PILOT for vulnerability detection in software systems. The lack of labeled negative examples makes traditional supervised learning impractical in this field, but PU learning can effectively utilize the available positive and unlabeled data to identify vulnerabilities that have gone unnoticed, such as the critical CVE-2007-4559 vulnerability in Python’s tarfile module. 
        
        \textit{Wu et al.} \cite{Wu2019} explored the use of PU learning for identifying compromised hosts in a network. In their study, the majority of network traffic is unlabeled, and only a small portion of the data is labeled as risky. They proposed a method where positive labels are propagated throughout the dataset, transforming the problem into a binary classification task that can detect previously unidentified compromised hosts.
        
        For malware detection, \textit{Fan et al.} \cite{Fan2023} propose a PU learning framework to detect malicious traffic in network communications. This method helps overcome the challenges of class imbalance and insufficient labeled data, which are common issues in malware detection. By leveraging PU learning, Fan et al. were able to identify new forms of malicious traffic from a predominantly benign dataset, improving both detection accuracy and response time.

    \subsection{PU Learning for Anomaly Detection and URL Attack Detection}
    
        Anomaly detection in cybersecurity is crucial for identifying suspicious behaviors that could signify security breaches. Traditional supervised methods require large labeled datasets to detect such anomalies, but PU learning can operate effectively with limited labeled data and abundant unlabeled data.
        
        \textit{Ju et al.} \cite{Ju2020} demonstrate the application of PU learning in anomaly detection tasks. By employing deep metric learning and a novel filtering method, their model successfully identifies anomalous patterns in unlabeled datasets. This method could be applied to detect anomalies in network traffic, where the majority of data is normal, and only a few instances represent cyber threats.
        
        In a similar vein, \textit{Zhang et al.} \cite{Zhang2017} explore the use of PU learning for detecting URL-based attacks. Traditional URL attack detection systems rely on blacklists or supervised machine learning, both of which require labeled negative samples. However, in the real world, it is challenging to label all malicious URLs. Zhang et al. formalize the URL detection problem as a PU learning task, where known malicious URLs serve as positive samples, and the rest remain unlabeled. This method provides better generalization in detecting newly generated URL attacks that might evade traditional blacklists.
    
    \subsection{Applications of PU Learning in DNS and Domain Name Security}
    
        Domain Name System (DNS) security is another important area in cybersecurity. Malicious domains are often used for phishing, botnet communication, and malware distribution. Detecting these domains is crucial for stopping cyber-attacks early.
        
        \textit{Fan et al.} \cite{Fan2022} propose a PU learning framework for detecting malicious domains by analyzing DNS traffic. Given the high cost of manual labeling, their model only uses a small set of labeled malicious domains combined with a large set of unlabeled DNS traffic to build an effective classifier. This approach helps alleviate the problem of class imbalance, where benign domains outnumber malicious ones. By customizing sample weights in the loss function, their PU-based model improves detection accuracy and reduces false positives, enhancing domain security.
        
        \subsection{Meta-Learning in PU Classification}
        PU learning has also gained attention in meta-learning, where the goal is to generalize learning across multiple tasks. In cybersecurity, where threats constantly evolve, being able to adapt learning models to new threats without retraining on vast amounts of data is critical.
        
        \textit{Kumagai et al.} \cite{Kumagai2024} introduce a meta-learning approach for PU classification that uses knowledge from related tasks (with positive, negative, and unlabeled data) to enhance performance on tasks with only positive and unlabeled data. For instance, in network security, models trained on data from previous intrusion detection tasks can be adapted to detect novel attacks in new environments. This meta-learning method minimizes the classification risk and improves accuracy, even in scenarios with limited labeled data.
    
    \subsection{Challenges in Labeled Data and Density-Ratio Estimation}
    
        Collecting large amounts of labeled data remains a significant bottleneck in many real-world machine learning applications, including cybersecurity. Labeled negative examples are often difficult to gather due to the dynamic and diverse nature of cyber threats.
        
        \textit{Sakai et al.} \cite{Sakai2016} argue that the scarcity of labeled data can be mitigated using PU learning, as it allows classifiers to be trained using only positive and unlabeled data. They propose the use of unbiased risk estimators that leverage unlabeled data for risk evaluation, without requiring restrictive distributional assumptions. This approach is particularly relevant in cybersecurity applications like vulnerability detection and malware identification, where the negative class is often underrepresented or mislabeled.
        
        \textit{Wang et al.} \cite{Wang2023} further explore PU learning in situations where labeled data is scarce and costly to obtain. They show that complementary-label learning can be expressed as a negative-unlabeled classification problem, and propose a model that adapts to the limitations of real-world cybersecurity datasets by estimating the density ratio of PU data. By using a density-ratio estimation method, their model effectively improves classification accuracy in domains where class priors are unknown or difficult to estimate.
    
    \subsection{Generative Adversarial Networks and PU Learning}
    
        The integration of Generative Adversarial Networks (GANs) with PU learning has shown promise in scenarios where class labels are ambiguous or difficult to obtain. GANs have been widely applied in cybersecurity for tasks such as malware detection and traffic anomaly detection.
        
        \textit{Zhang et al.} \cite{Zhang2024} introduce Predictive Adversarial Networks (PANs), which extend GANs by replacing the generator with a classifier for PU learning tasks. In their framework, the classifier aims to predict the positive class from the unlabeled data, while the discriminator differentiates between real positive samples and predicted positives. This approach is highly relevant in domains like malware detection, where it is difficult to acquire large labeled datasets. By leveraging adversarial learning, PANs improve the classifier's ability to distinguish between benign and malicious instances in cybersecurity datasets.
    
    \subsection{Cost-Sensitive Learning and Label Disambiguation in PU Learning}
    
        One of the major challenges in PU learning is label ambiguity. In cybersecurity, this often arises due to the difficulty of distinguishing between benign and malicious activities, especially in real-time environments like intrusion detection systems.
        
        \textit{Long et al.} \cite{Long2024} propose a cost-sensitive PU learning framework that addresses label disambiguation through importance reweighting and meta-learning. In their method, reliable negative and positive instances are selected from the unlabeled set, improving the overall quality of the training data. This is particularly useful in dynamic environments such as security operations centers (SOCs), where detecting high-risk threats among a large volume of benign traffic is a pressing challenge. By reducing the reliance on unreliable labels, their approach enhances the robustness of PU models in cybersecurity applications.

    \subsection{Challenges and Future Directions}
        Although PU and NU learning hold significant potential, several challenges remain. One key issue, as noted by \textit{Foong et al.} \cite{Foong2019}, is the difficulty in obtaining high-quality labeled data. In many cases, positive samples are scarce and expensive to collect, while negative samples are diverse and difficult to accurately label. 
        
        \textit{Le et al.} \cite{Le2020} further highlight the challenge of retrieving security-related content from Q\&A websites using PU learning, where obtaining reliable negative samples is impractical. They suggest that future research should focus on improving methods for identifying reliable negative samples in PU learning, which would significantly enhance its applicability in cybersecurity.
        
        Another challenge is the computational complexity associated with real-time detection. \textit{Jaskie et al.} \cite{Jaskie2019} noted that while PU learning models are effective, they may not scale efficiently for large-scale, real-time systems like Intrusion Detection Systems. Future research should aim to optimize PU learning algorithms to handle large volumes of streaming data in real-time environments.

    \subsection{Key Takeaways}
        
        The literature demonstrates that PU and NU Learning are well-suited for addressing the challenges of limited labeled data in cybersecurity. From detecting vulnerabilities and malware to identifying phishing attacks and compromised hosts, these methods offer a robust framework for leveraging unlabeled data. However, challenges such as computational efficiency and data quality remain, providing ample opportunity for further research and development.

\section{Research Gaps}

    Although Positive Unlabeled (PU) Learning and Negative Unlabeled (NU) Learning have been successfully applied to a range of cybersecurity tasks, there are several critical research gaps that must be addressed to unlock their full potential. These gaps arise from both the complexity of cybersecurity challenges and the limitations of existing PU/NU learning techniques.
    
    \subsection{Scalability in Real-Time Systems}
    
        Many cybersecurity systems, such as Intrusion Detection Systems (IDS) and Security Information and Event Management (SIEM) systems, operate in real-time, monitoring vast volumes of data. Current PU and NU learning algorithms struggle with the computational overhead required to process such data at scale. \textit{Fan et al.} \cite{Fan2022} demonstrated how PU learning can be applied to DNS security, but the framework lacks the scalability to handle real-time DNS traffic in large-scale networks. There is a need for lightweight, scalable PU/NU models that can be integrated into real-time systems without compromising speed or accuracy.
    
    \subsection{Limited Labeled Data for Emerging Threats}
    
        While PU learning is highly effective in scenarios where labeled data is scarce, it still relies on the availability of some labeled positive examples. In many cybersecurity domains, such as malware detection and vulnerability management, new types of attacks or vulnerabilities emerge regularly. As noted by \textit{Lv et al.} \cite{Lv2020}, these new threats are difficult to address due to the lack of labeled data, especially in the initial stages of attack detection. This creates a gap in developing adaptive PU/NU learning algorithms that can quickly adapt to unseen threats with minimal supervision.
    
    \subsection{Class Imbalance and Label Ambiguity}
    
        Cybersecurity datasets are often highly imbalanced, with benign activities vastly outnumbering malicious ones. \textit{Long et al.} \cite{Long2024} highlight that label ambiguity--common in cybersecurity datasets where the distinction between benign and malicious activities is unclear--poses a significant challenge for PU learning. Addressing class imbalance and label disambiguation remains a research gap, particularly in contexts like security operations, where noisy or unreliable labels are frequent.
    
    \subsection{Incorporating Domain Knowledge}
        
        Although PU/NU learning provides flexibility in working with limited labeled data, cybersecurity tasks often require incorporating specific domain knowledge (e.g., known attack patterns or cryptographic properties). Current PU/NU learning models do not effectively leverage domain-specific insights. As seen in \textit{Wang et al.} \cite{Wang2023}, there is potential to enhance classification accuracy by integrating domain knowledge into the PU learning framework, such as by adjusting risk estimators or incorporating expert-annotated features.
    
    \subsection{Generalization Across Cybersecurity Subfields}
    
        While PU and NU learning have shown promise in specific subfields like anomaly detection and malware identification, their generalization across diverse cybersecurity subfields is still limited. For example, \textit{Kumagai et al.} \cite{Kumagai2024} explore meta-learning in PU classification to generalize across tasks, but such methodologies have not been widely applied in other areas like cryptography or risk management. A broader understanding of how these learning techniques can be tailored for different cybersecurity domains--such as cryptography or IoT security--remains underexplored.
    
    \subsection{Handling Dynamic and Evolving Threats}
    
        Many cybersecurity tasks involve detecting threats that evolve over time, such as Advanced Persistent Threats (APTs) or zero-day vulnerabilities. PU and NU learning models are generally static and trained on historical data, making them less effective in responding to rapidly evolving threats. While \textit{Zhang et al.} \cite{Zhang2024} propose using adversarial networks for PU learning in dynamic scenarios, there remains a gap in designing adaptive learning algorithms that can continuously update as new threats emerge.

\section{Problem Formulation}

    To address these research gaps, we propose a formal problem formulation for applying PU and NU learning to key cybersecurity challenges. Each cybersecurity subfield can be framed as a classification problem involving positive, negative, and unlabeled data, depending on the availability of labeled examples. Below, we outline the formulation of PU/NU learning in several critical areas of cybersecurity, linking them to the challenges discussed earlier.
    
    \subsection{General Framework for PU Learning}
    
        In PU learning, the goal is to classify data points into positive (malicious) or negative (benign) classes using a dataset where only positive examples \( P = \{x_i | y_i = 1\} \) are labeled, and the remaining samples are unlabeled \( U = \{x_i | y_i = ?\} \). The objective is to minimize the classification risk:
        \begin{equation}
        R(f) = \mathbb{E}_{x \in P}[\ell(f(x), 1)] + \mathbb{E}_{x \in U}[\ell(f(x), -1)]
        \end{equation}
        where \( \ell(f(x), y) \) is a loss function, and \( f(x) \) is the classifier. The challenge in cybersecurity is to effectively estimate this risk in the context of highly imbalanced and evolving data.
    
    \subsection{Network Security and Intrusion Detection}
    
        Network security involves classifying network traffic as benign or malicious. Given the imbalance between the number of benign (negative) and malicious (positive) events, this can be formulated as a PU learning problem. Let network traffic \( T = \{t_1, t_2, \dots, t_n\} \) be the dataset, where only malicious traffic \( P \) is labeled. The task is to detect malicious traffic from the unlabeled dataset \( U \), which contains mostly benign traffic.
        
        We can model the problem as minimizing a classification risk \( R(f) \) similar to the general PU learning framework, with adjustments to account for class imbalance through cost-sensitive learning or sample reweighting:
        \begin{equation}
        R(f) = \mathbb{E}_{x \in P}[\ell(f(x), 1)] + \lambda \mathbb{E}_{x \in U}[\ell(f(x), -1)]
        \end{equation}
        where \( \lambda \) is a weight factor that adjusts for the imbalance between the positive and unlabeled sets.
    
    \subsection{Vulnerability Management and Application Security}
    
        In vulnerability management, the goal is to identify security vulnerabilities in software code. Let \( C = \{c_1, c_2, \dots, c_n\} \) represent code snippets from a software application, with \( V = \{v_1, v_2, \dots, v_k\} \) representing known vulnerable code fragments. The problem can be framed as a PU learning task, where the known vulnerabilities form the positive set \( P \), and the rest of the code remains unlabeled \( U \). The goal is to predict which unlabeled code fragments are likely to contain vulnerabilities, minimizing the classification risk.
        
        NU learning could also be applied when there is a substantial set of benign code \( N \), and the goal is to identify potential vulnerabilities \( V \) in a mostly benign codebase.
    
    \subsection{Threat Intelligence and Malware Detection}
    
        Threat intelligence aims to detect new, previously unseen malware samples from a large dataset of benign files. Let \( M = \{m_1, m_2, \dots, m_n\} \) represent a set of files, with a small set \( P \) of labeled malware files. The challenge is to identify additional malware from the unlabeled set \( U \) of files. This can be framed as a PU learning problem, where the classifier must distinguish between benign and malicious files using limited positive examples. Given the high dimensionality of the malware dataset, this task requires efficient feature extraction and representation learning to improve classification accuracy.
    
    \subsection{Incident Response and SIEM}
    
        In incident response, security events are classified as high-risk or low-risk based on alerts generated by a SIEM system. Let \( E = \{e_1, e_2, \dots, e_n\} \) represent security events, where only a few high-risk events \( P \) are labeled. The majority of events \( U \) are unlabeled, as security analysts do not have the resources to investigate every alert. The task is to predict the likelihood of a given event being high-risk based on its features, such as the type of attack or the severity of the alert. PU learning can be used to prioritize the investigation of high-risk events by learning from the small set of known positives.
    
    \subsection{IoT Security}
    
        In Internet of Things (IoT) security, the goal is to detect abnormal behaviors in a network of connected devices. Let \( D = \{d_1, d_2, \dots, d_n\} \) represent the IoT devices in a system, with a small set of known compromised devices \( P \). The majority of devices \( U \) are unlabeled, as their security status is unknown. The task is to predict which devices in \( U \) are potentially compromised, based on their behavioral patterns. Given the diversity of IoT devices and the dynamic nature of the data they generate, PU learning is well-suited for this task, allowing for efficient anomaly detection without requiring extensive labeled data.

\section{Conclusion}

    How can we secure an ever-evolving digital world where threats continuously change, data labeling is expensive, and traditional supervised learning struggles to keep pace? Positive Unlabeled (PU) Learning and Negative Unlabeled (NU) Learning offer promising solutions by efficiently handling large, mostly unlabeled datasets--common in cybersecurity. By leveraging these semi-supervised learning methods, we can make significant strides in mitigating the limitations of traditional approaches. But where exactly do PU/NU Learning shine, and what challenges do they overcome?
    
    One analogy to understand PU/NU Learning's role in cybersecurity is to think of cybersecurity as a vast, dark forest. In this forest, positive examples (malicious events) are rare but dangerous--like venomous creatures hiding in the shadows. On the other hand, the forest is teeming with harmless creatures (benign events) that go largely unnoticed. Traditional methods require tagging each creature before distinguishing the dangerous ones, but PU/NU Learning instead equips us with a map, showing where we know threats exist (positive examples) while helping us infer where other unseen dangers may be lurking (unlabeled data). This approach saves both time and resources, addressing key cybersecurity challenges.
    
    \subsection{What makes PU/NU Learning particularly suited to cybersecurity applications?}
    
        As \textit{Niu} \cite{Niu2017} notes, labeled data in cybersecurity is often either too expensive to obtain or too diverse and impure to be useful. For instance, obtaining reliable negative examples can be financially prohibitive or logistically impractical. PU Learning becomes essential in these scenarios, where positive examples (e.g., known attacks) are limited but unlabeled data (e.g., network traffic or event logs) is abundant. PU Learning excels because it does not require explicit negative labels, allowing models to learn from scarce positive examples and the surrounding unlabeled data.
        
        Consider the problem of incident response in a Security Information and Event Management (SIEM) system. \textit{Feng et al.} \cite{Feng2017} describe the overwhelming number of alerts generated by SIEM systems--far beyond the investigative capacity of security analysts. Here, PU learning can dramatically reduce false positives by identifying high-risk hosts with minimal labeled data. In this context, PU Learning acts as a sieve, filtering through vast amounts of alert data, allowing analysts to focus on truly critical incidents without needing an exhaustive set of labeled examples.
    
    \subsection{What are the limitations, and how can they be overcome?}
    
        The scalability and adaptability of PU/NU Learning remain critical challenges. As discussed by \textit{Fan et al.} \cite{Fan2022} and \textit{Long et al.} \cite{Long2024}, cybersecurity often involves large, imbalanced datasets. Furthermore, emerging threats and new attack patterns pose the problem of dynamic adaptation--current models, while effective, are largely static. PU/NU Learning methods need to evolve to handle real-time data streams and continuously adapt to new types of attacks.
        
        Additionally, handling label ambiguity and class imbalance, as highlighted by \textit{Long et al.}, is a persistent problem in cybersecurity applications. In this field, label uncertainty is not just an inconvenience; it can mean the difference between preventing an attack or suffering a costly breach. Solutions such as cost-sensitive learning and label disambiguation, integrated into the PU/NU framework, can help address these issues by assigning higher weights to critical alerts or adapting the decision boundaries between positive and unlabeled examples.
    
    \subsection{What is next for PU/NU Learning in cybersecurity?}
    
        The future of PU/NU Learning\footnote{A Python implementation of PU/NU learning can be found here \href{https://pulearn.github.io/pulearn/doc/pulearn/}{\texttt{pulearn}}.} in cybersecurity lies in tackling the complexities of diverse and evolving cyber threats. As we move forward, interdisciplinary efforts that incorporate domain knowledge from cybersecurity experts into learning models, as suggested by \textit{Wang et al.} \cite{Wang2023}, will be crucial for improving accuracy and efficiency. Additionally, extending PU/NU Learning to more specialized areas like IoT security, cryptography, and compliance will require further refinement of these models.
    
    To summarize, PU and NU Learning have the potential to transform cybersecurity by addressing the challenges of limited labeled data and scaling to massive, dynamic environments. They allow us to navigate the \say{dark forest} of cybersecurity with greater efficiency, finding hidden threats that would otherwise go unnoticed. The journey ahead involves refining these models for real-time applications, improving scalability, and integrating domain expertise to continue adapting to the ever-changing threat landscape.

\bibliographystyle{ACM-Reference-Format}
\bibliography{references}

\end{document}